# Ferromagnetism in orthorhombic RAgAl$_3$ (R = Ce and Pr) compounds

S. Nallamuthu[1], Andrea Dzubinska[2], Marian Reiffers[2], Jesus Rodriguez Fernandez[3] and R. Nagalakshmi[1*]

[1]Department of Physics, National Institute of Technology, Tiruchirappalli 620 0015, Tamil Nadu, India

[2]Faculty of Humanities and Natural Sciences, Presov University, Presov, Slovakia

[3]Departamento CITIMAC, Facultad de Ciencias, Universidad de Cantabria, Santander, Spain

## Abstract

We present a detailed study on magnetic, thermodynamic and transport properties of polycrystalline RAgAl$_3$(R = Ce and Pr) compounds. Both compounds crystallize in orthorhombic structure, which is distorted from the tetragonal BaAl$_4$ structure with the space group *Cmcm*. Heat capacity measurement indicates the bulk magnetic ordering of the compounds. CeAgAl$_3$ and PrAgAl$_3$ order ferromagnetically at T$_C$ = 3.8 K and 5.8 K, respectively as it was confirmed from magnetic measurements. CeAgAl$_3$ exhibits heavy Fermion behaviour. The Schottky behaviour in heat capacity data was observed in both compounds. The crystalline electric field (CEF) analysis of the magnetic parts of heat capacity of CeAgAl$_3$ and PrAgAl$_3$ yielded to a CEF level scheme with three doublets and nine singlets and with an overall splitting of 51 K and 180 K, respectively. Fit yielded a magnetic doublet state for CeAgAl$_3$, whereas for PrAgAl$_3$ a pseudo-doublet ground-state with an energy difference of 15 K has been obtained. The resistivity measurements display a low temperature drop at the magnetic ordering temperature of the compounds. Negative magnetoresistance (MR) due to the ferromagnetic ordering has been observed for both Ce and Pr compounds.

*Corresponding author: Tel.: 04312503615; e-mail: nagaphys@yahoomail.com, nagalakshmi@nitt.edu (R.Nagalakshmi),

## 1. Introduction

Rare earth intermetallic compounds often show a variety of interesting physical properties due to the competition of electronic interactions between localized f-electrons and itinerant d-electrons. The primary interaction occurs via the polarization of the conduction electrons, which is known as RKKY (Ruderman-Kittel-Kasuya-Yosida) interactions (indirect exchange) and the 4f electrons are strongly affected by their local environment. CEF is the factor contributing to large variety of magnetic, thermodynamic and electronic properties in rare earth intermetallics. Different centrosymmetric and non-centrosymmetric structures have been crystallized for various combinations of the RTX$_3$ (R = rare earth, T = 3$d$/4$d$/5$d$ - transition metal and X = p-block element)-type crystal structure. Recently reported several RTX$_3$ compounds, crystallizing in BaNiSn$_3$ type structure, exhibit a huge range of unusual magnetic or superconducting ground state properties. Of which, CePt$_3$Si [1] and CeRhSi$_3$ [2, 3] have been receiving much attention due to the existence of spin orbit coupling induced by non-inversion symmetry. Besides the tetragonal BaNiSn$_3$ structure the existence of new ternary compounds RNi$_x$Ga$_{4-x}$ with CePtGa$_3$ structure have also been reported [4]. At the composition RPtGa$_3$, within the homogeneous range of the BaAl$_4$ type structure, the existence of the compound with an orthorhombic deviation from BaAl$_4$ was observed [5, 6]. In continuation of our studies on RTX$_3$ compounds such as RCuGa$_3$(R = Pr, Nd and Gd) [7] and RCoSi$_3$ (R = Pr, Nd and Sm)[8], we have undertaken the systematic studies of polycrystalline RAgAl$_3$ (R = La, Ce and Pr) samples possessing orthorhombic crystal structure. Initial magnetic and thermodynamic properties of ferromagnetic CeAgAl$_3$ single crystals have been reported [9]. First report on single crystalline CeAgAl$_3$ is found to exhibit ferromagnetic (T$_C$ = 3K) character suggesting BaNiSn$_3$ type structure (space group = *I4/mm*) or PrNiGa$_3$ structure (space group = *I4/mmm*) [9]. Subsequently Franz et. al solved the structural controversy of single crystalline CeAgAl$_3$ and reported to be orthorhombic structure with the space group *Cmcm* (No. 63) [10]. Herein, we report the systematic investigations of structural, magnetic, thermodynamic, transport and magneto-transport properties of orthorhombic RAgAl$_3$ (Ce and Pr) polycrystalline compounds for the first time.

## 2. Experimental methods

The polycrystalline samples of RAgAl$_3$ (R = La, Ce and Pr) were synthesized by arc-melting under a purified argon atmosphere. The elements of high purity (La (99.9%), Ce (99.8%), Pr (99.9%), Ag (99.9999%), Al (99.999%)) were taken in the stoichiometric 1:1:3 ratios. The resulting alloy button was turned over and remelted several times to ensure the homogeneity. The ingots were annealed at 700ºC under vacuum in a quartz ampoule for 20 days. To confirm the phase purity of the annealed sample, powder X- ray diffraction using Cu-Kα was carried out on the annealed sample. Stoichiometry and single phase nature were determined by energy-dispersive X-ray spectroscopy (EDAX) measurements.

dc magnetization was measured as a function of temperature and magnetic field [isothermal magnetization] using the Quantum Design, Physical Property Measurement System (PPMS) from 2 K to 300 K up to the field of 50 kOe. In the zero field cooled (ZFC) process, the sample is cooled below $T_C$ in the absence of the magnetic field down to 2 K and then the magnetization was measured with increasing the temperature in constant applied magnetic field. In the field cooled (FC) process, the measurements were done in the applied magnetic field while the sample was cooling down to 2K. The ac susceptibility measurements were done as a function of temperature at various frequencies from 10 to 1488 Hz in a probing field of 2.5 Oe using the magnetic properties measurements systems (MPMS). Heat capacity was also measured using PPMS relaxation technique down to 1.8 K with external magnetic field up to 90 kOe. Also, four-probe resistivity and magnetoresistance (MR) measurements were carried out using PPMS from 1.8 to 300 K in zero field and magnetic fields up to 80 kOe.

## 3. Results
## 3.1. X-ray diffraction

Fig.1. shows the powder X-ray diffraction pattern of $CeAgAl_3$ (recorded at room temperature) along with structural Rietveld refinement profile using GSAS software [11]. From the results obtained by Rietveld refinement of $RAgAl_3$ (R = La, Ce and Pr), it is confirmed that the compounds adopt orthorhombic $SrPdGa_3$ - type structure (space group *Cmcm*) revealing a distortion from the tetragonal $BaAl_4$-type parent structure [12]. The crystallographic information related to lattice parameters and atomic positions are given in the Table 1 and Table 2, respectively. Also, a small Ag/Al1 mixing occurs at the 4c sites, which are in good agreement with the recent report on single crystals of $CeAgAl_3$ [10].

**Table 1**. Rietveld refinement results of RAgAl$_3$ (R = La, Ce and Pr) having the orthorhombic structure.

| Compound | $a$ (Å) | $b$ (Å) | $c$ (Å) | $V_{cell}$ (Å$^3$) | $\chi^2$ | $R_p$ (%) | $R_{wp}$ (%) |
|---|---|---|---|---|---|---|---|
| LaAgAl$_3$ | 6.1515(19) | 10.8666(29) | 6.2512(62) | 418.46(8) | 1.7 | 21 | 28 |
| CeAgAl$_3$ | 6.2227(2) | 10.8383(92) | 6.1149(59) | 412.76(2) | 2.1 | 15 | 21 |
| PrAgAl$_3$ | 6.1874 (4) | 10.8400 (99) | 6.0834(82) | 408.03(2) | 6.3 | 19 | 31 |

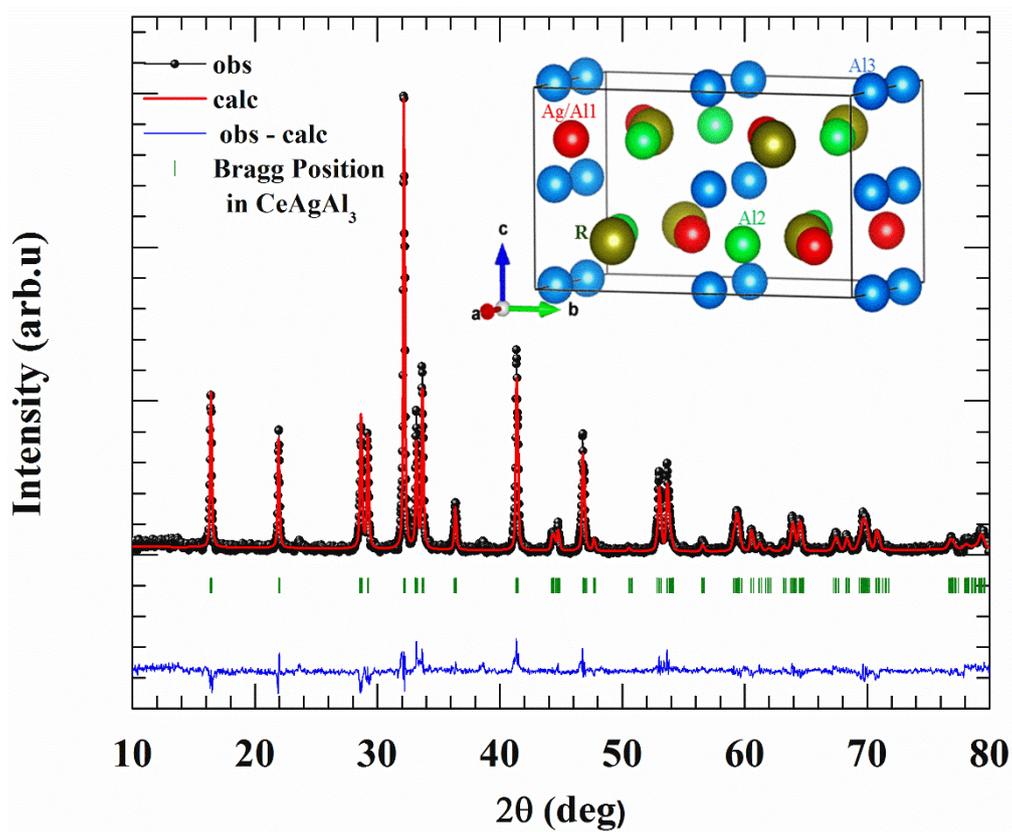

**Fig.1**. Powder X-ray diffraction pattern of orthorhombic CeAgAl$_3$ along with Rietveld refinement using GSAS program. The inset shows the crystal structure of RAgAl$_3$ (R = La, Ce and Pr). The atoms are represented by different colours and ball size. The dark line shows the unit cell.

**Table 2**. Atomic positions obtained from Rietveld refinement of RAgAl$_3$ (R = La, Ce and Pr)

| Atom | Site | x | y | z | Occupancy |
|---|---|---|---|---|---|
| **LaAgAl$_3$** | | | | | |
| La | 4c | 0.0000 | 0.2423(97) | 0.2500 | 1.00 |
| Ag | 4c | 0.5000 | 0.3780(54) | 0.2500 | 0.972(7) |
| Al1 | 4c | 0.5000 | 0.3780(54) | 0.2500 | 0.028(7) |
| Al2 | 4c | 0.5000 | 0.1264(87) | 0.2500 | 1.00 |
| Al3 | 8e | 0.2231(29) | 0.5000 | 0.0000 | 1.00 |
| **CeAgAl$_3$** | | | | | |
| Ce | 4c | 0.0000 | 0.2465(30) | 0.2500 | 1.00 |
| Ag | 4c | 0.5000 | 0.3851(19) | 0.2500 | 0.928(2) |
| Al1 | 4c | 0.5000 | 0.3851(19) | 0.2500 | 0.072(8) |
| Al2 | 4c | 0.5000 | 0.1628(33) | 0.2500 | 1.00 |
| Al3 | 8e | 0.2297(19) | 0.5000 | 0.0000 | 1.00 |
| **PrAgAl$_3$** | | | | | |
| Pr | 4c | 0.0000 | 0.2434(53) | 0.2500 | 1.00 |
| Ag | 4c | 0.5000 | 0.3840(14) | 0.2500 | 0.898(8) |
| Al1 | 4c | 0.5000 | 0.3840(14) | 0.2500 | 0.102(6) |
| Al2 | 4c | 0.5000 | 0.1442(33) | 0.2500 | 1.00 |
| Al3 | 8e | 0.2365(81) | 0.5000 | 0.0000 | 1.00 |

## 3.2. Magnetic Susceptibility and Magnetization

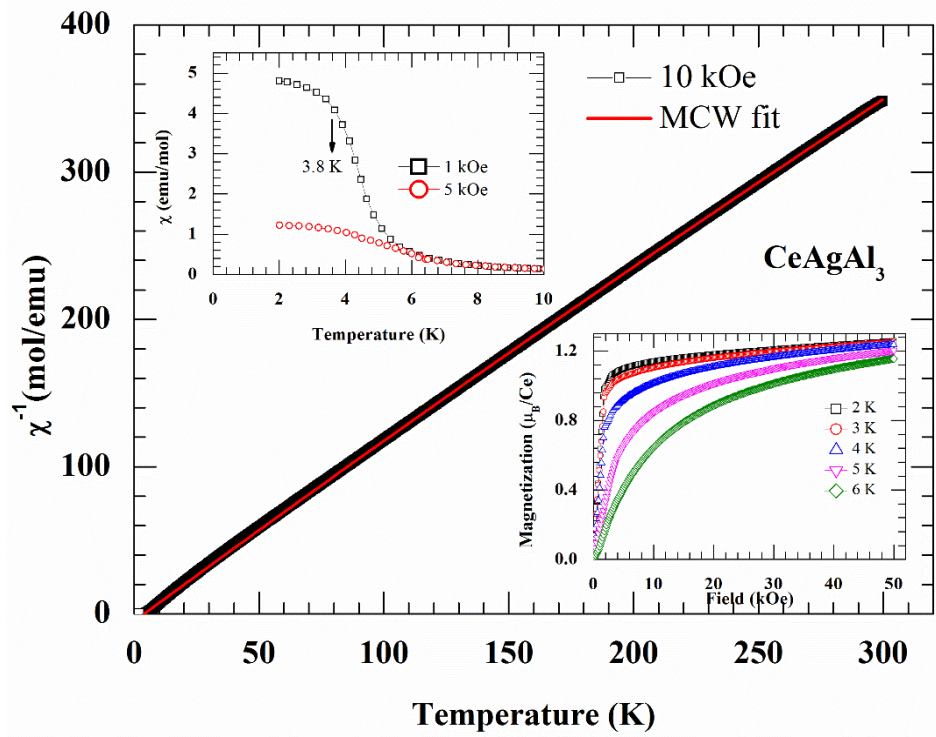

**Fig.2.** Inverse magnetic susceptibility of CeAgAl3. The upper inset shows susceptibility ($\chi$) in 1 and 5 kOe below 10 K. The lower inset shows isothermal magnetization versus applied field at selected temperatures.

The inverse magnetic susceptibility ($\chi^{-1}$) of CeAgAl3 measured as a function of temperature from 2 - 300 K in an applied field of 1 kOe is shown in Fig. 2. Magnetic susceptibility of CeAgAl3 exhibits a ferromagnetic transition with $T_C$ = 3.8 K (inset of Fig.2), which is in agreement with the previously reported result [9]. The linear paramagnetic region (100 – 300 K) has been fitted with the modified Curie-Weiss (MCW) formula $\chi = \chi_0 + C/(T - \theta_P)$ ($\chi_0$ is the temperature-independent term, attributed to the contributions of conduction electrons and $C$ is Curie constant ($C = \mu_{eff}^2 x/8$) ). The estimated parameters $\chi_0 = 1.501\times10^{-4}$ emu/mol, $\theta_p$ = 4 K and effective magnetic moment $\mu_{eff}$ = 2.53 $\mu_B$/Ce. Where $\mu_{eff}$ is in good agreement with expected theoretical value and it indicates the trivalent state of $Ce^{3+}$ free ion (2.54 $\mu_B$). The positive $\theta_p$ indicates ferromagnetic correlation at low temperatures. The magnetic isotherm of CeAgAl3 up to 50 kOe is shown in the lower inset of Fig. 2. The magnetization increases rapidly below 2 kOe proposing to be ferromagnetic and tends to attain the magnetization value of 1.26 $\mu_B$/Ce at 50 kOe. This value is lower than the saturated moment for the ground-state multiplet of the free ion Ce$^{3+}$ ($g_J J\mu_B$ = 2.14 $\mu_B$/Ce), which suggests the prevalence of the crystal electric field effect in this compound.

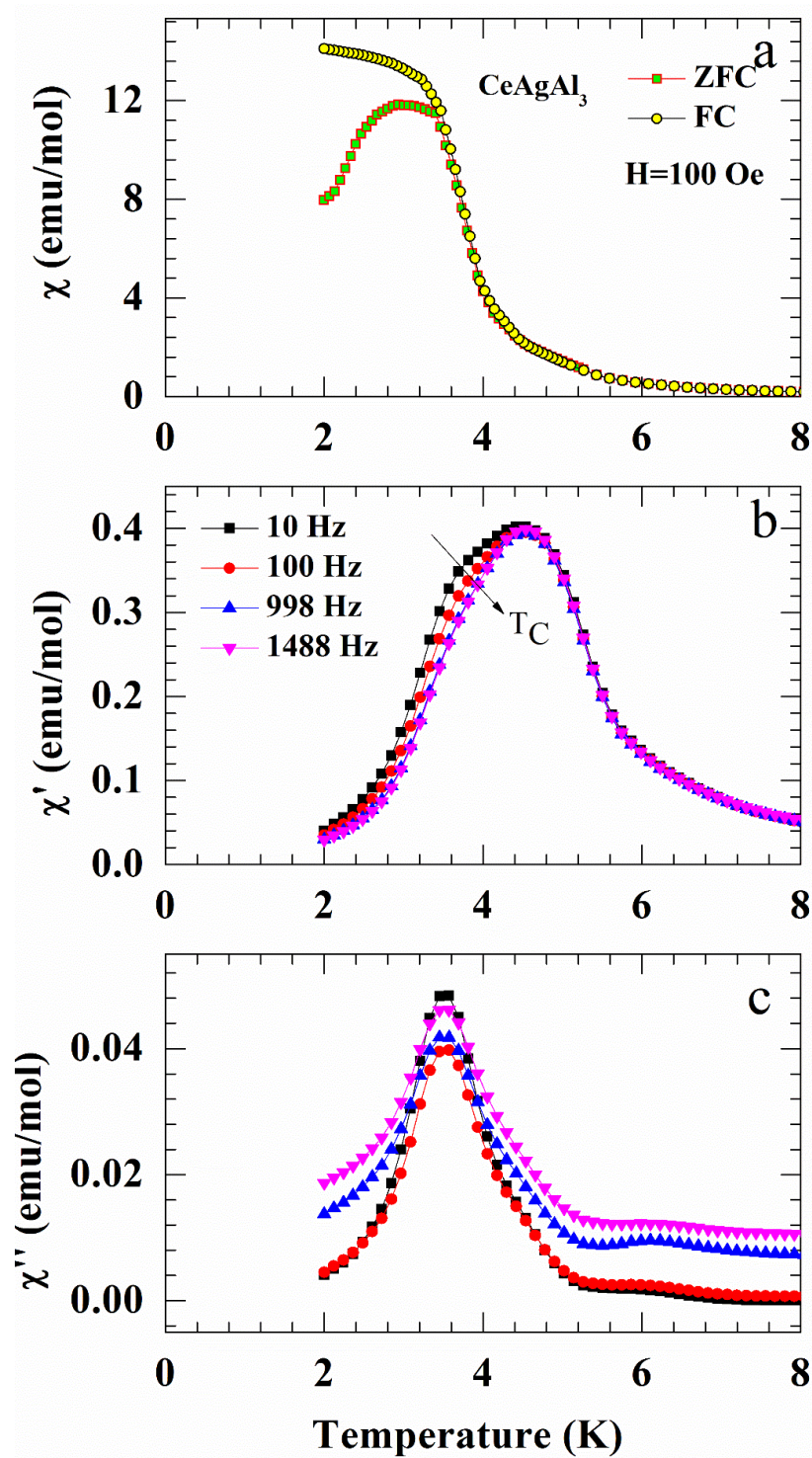

**Fig.3**. dc and ac magnetic susceptibility of CeAgAl$_3$. (a). dc magnetic susceptibility measured under ZFC and FC conditions. (b). Real part $\chi'$ and (c) Imaginary part $\chi''$ of ac susceptibility measured at various frequencies such as 10, 100, 998, 1488 Hz for $H_{ac}$ = 2.5 Oe.

In order to investigate the magnetic properties of CeAgAl₃, the temperature dependence of magnetization was carried out in ZFC and FC modes at $H = 100$ Oe, as shown in Fig.3a. There is a slope change at ferromagnetic Curie temperature $T_C = 3.8$ K. However, there is a difference between ZFC and FC curves indicating thermomagnetic irreversibility. This strong irreversibility of ZFC and FC magnetization curves below $T_{irr}$ is expected to disappear as applied field is increased. This irreversibility behaviour is typically attributed either to domain wall pinning effect of ferromagnetic domains [13] or the presence of spin glass behaviour [14]. Further ac magnetic susceptibility measurements were carried out to confirm the spin glass behaviour. We measured the temperature dependence of the ac susceptibility for various frequencies such as 10, 100, 998 and 1488 Hz. The real and imaginary parts of the ac magnetic susceptibilities $\chi'$ and $\chi''$ are depicted Fig.3. The ac susceptibility did not show any prominent frequency dependence at the ordering temperature which rules out the possibility of possessing spin glass nature. Hence this suggests that the bifurcation between ZFC and FC dc susceptibility in CeAgAl₃is completely ascribed to the existence of ferromagnetic domain wall pinning.

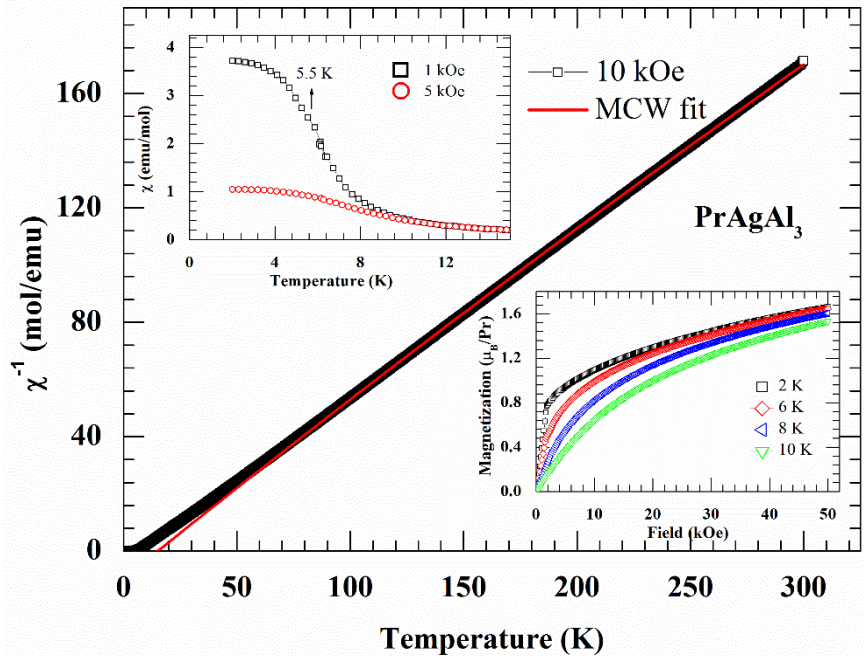

**Fig. 4**. Inverse magnetic susceptibility of PrAgAl₃. The upper inset shows susceptibility ($\chi$) in 1 and 5 kOe below 15 K. The lower inset shows isothermal magnetization versus applied field for different temperatures.

PrAgAl$_3$ orders ferromagnetically at 5.5 K (see upper inset of Fig. 4), similar to CeAgAl$_3$. The magnetic susceptibility in the paramagnetic region was fitted with modified Curie-Weiss formula (main plot of Fig.4). The obtained parameters are $\chi_0$ = 3.5544×10$^{-4}$ emu/mol, $\theta_p$ = 15 K and $\mu_{eff}$ = 3.54 $\mu_B$/Pr. The obtained $\mu_{eff}$ value is close to the theoretical value of Pr$^{3+}$ (3.58 $\mu_B$) free ion. Positive $\theta_p$ indicates the presence of ferromagnetic correlations. Isothermal magnetization measured as a function of field up to 50 kOe for various temperatures are shown in the lower inset of Fig.4. There is a rapid increase in magnetization up to 1.8 kOe suggesting the ferromagnetic interactions. Above 2 kOe, it behaves linearly up to the maximum field of 50 kOe and attains a value of 1.67 $\mu_B$ at 50 kOe. This value is smaller than the saturated moment for the ground-state multiplet of the free ion Pr$^{3+}$ ($g_JJ\mu_B$ = 3.2 $\mu_B$/Pr), which can be attributed to the presence of CEF.

## 4. Heat Capacity

The temperature dependent specific heat of RAgAl$_3$ (R = Ce, Pr) and its isostructural nonmagnetic reference system LaAgAl$_3$ are shown in Fig. 5. The specific heat data of CeAgAl$_3$ manifest a clear, sharp anomaly at 3.8 K with a peak height of 11.74 J/mol K indicates the bulk magnetic ordering of the Ce$^{3+}$ ion, which is close to the mean field value of 12.5 J/mol K for spin ½. The magnetic 4$f$ contribution to the heat capacity ($C_{4f}$) of the compound was deduced by subtracting the polycrystalline nonmagnetic analogue of LaAgAl$_3$. The magnetic entropy ($S_{4f}$) estimated by integrating $C_{4f}/T$ is shown in Fig. 5. At the ordering temperature, 60 % of the $S_{4f}$ of CeAgAl$_3$ (3.83 J/mol K$^2$) is released and completely reaching $R$ln(2) at 6.5 K, This suggests the presence of well isolated doublet ground state for CeAgAl$_3$. $R$ln(4) of the $S_{4f}$ is recovered at 27 K, which is close to the energy difference between first and second excited doublets. Above T$_C$, $S_{4f}$ increases gradually and attaining the maximum value of 14.7 J/mol K$^2$ at 100 K comparable to $R$ln(6). Total specific heat ($C(T)$) of PrAgAl$_3$ shows a well-defined $\lambda$-shaped anomaly at 5.5 K due to the ferromagnetic order (Fig.5). The magnetic heat capacity of PrAgAl$_3$ is obtained from subtracting the non-magnetic analogue LaAgAl$_3$. The $S_{4f}$ of PrAgAl$_3$ increases with increasing temperature, there by reaching the maximum of Rln9 at 100 K. The Sommerfeld electronic coefficient $\gamma$ and Debye temperature ($\theta_D$) have been estimated by using the equations (1 and 2).

$$C_P = \gamma T + \beta T^3 + \delta T^5 \qquad (1)$$

Where $\gamma$ is the electronic coefficient above ordering temperature (also referred to paramagnetic Sommerfeld coefficient $\gamma_P$), $\beta$ is Debye $T^3$ law lattice heat capacity coefficient and $\delta$ is higher order lattice terms. The estimated parameters ($\gamma$, $\beta$, $\delta$) obtained from fitting of equation (1) in the paramagnetic region ($T > T_C$) are given in Table 3 and the fitting is shown in Fig.5. Debye temperature ($\theta_D$) has been calculated from $\beta$ using the relation

$$\theta_D = \left(\frac{12\pi^4 Rn}{5\beta}\right)^{1/3} \quad (2)$$

where $R$ is the molar gas constant and $n$ is the number of atoms per formula unit. The calculated values are given in Table 3. Also, the Sommerfeld electronic coefficient $\gamma$ determined by extrapolation of $C/T$ up to lowest temperatures for PrAgAl$_3$ gives the value of 15 mJ/molK$^2$. This indicates that PrAgAl$_3$ is a normal metal. For better determination, it is necessary to perform measurements below 2 K, where this magnetic contribution diminishes.

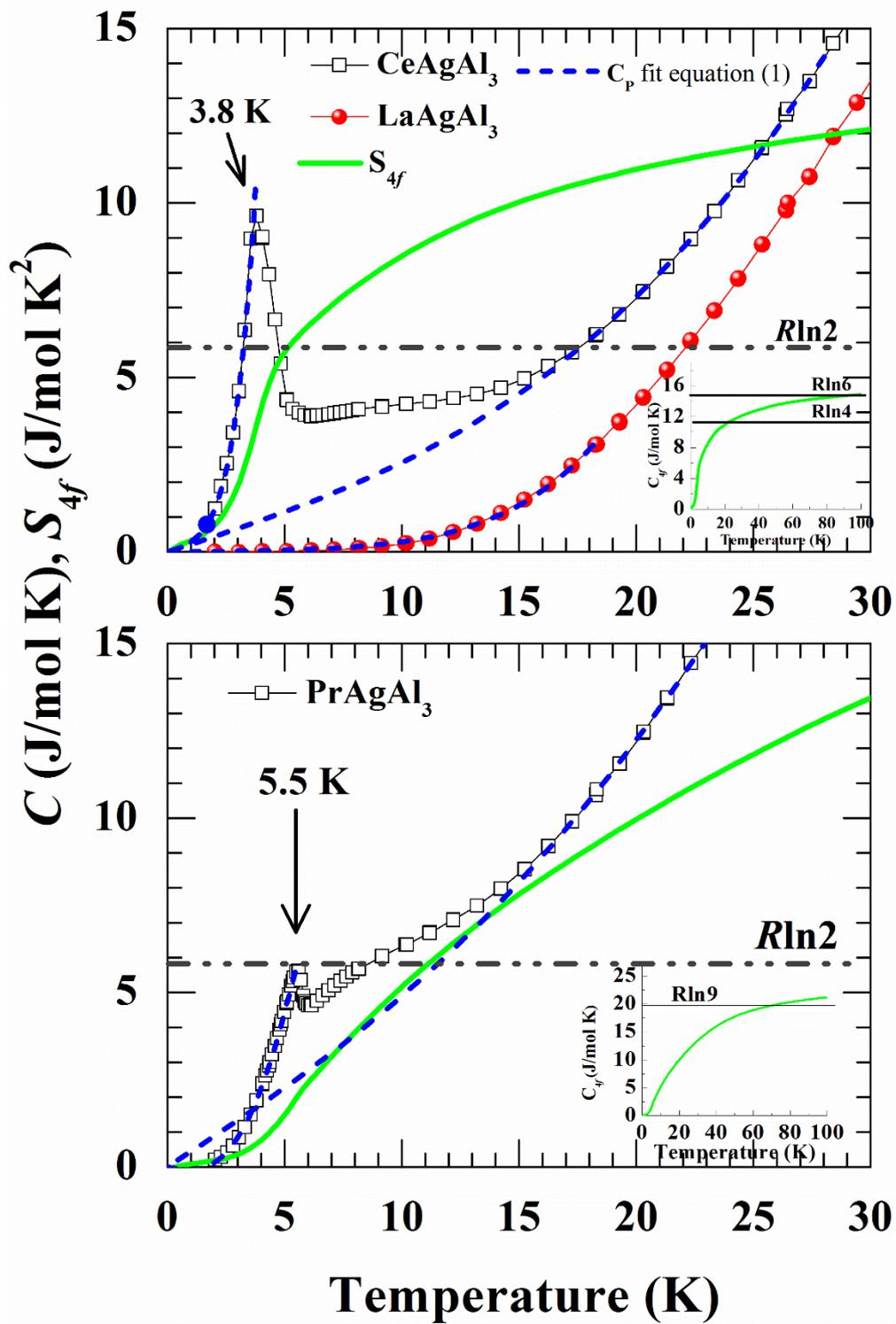

**Fig. 5**. Temperature dependent specific heat of RAgAl$_3$ (R = La, Ce and Pr) between 2 and 30 K. Dashed lines are fits to the equation $C_P = \gamma T + \beta T^3 + \delta T^5$. Insets of both figures show the maximum of magnetic entropy ($S_{4f}$) versus temperature.

**Table 3.** The parameters obtained from fitting the heat capacity data using equation (1) above the ordering temperature ($T > T_C$).

| Compounds | $\gamma$ (mJ/mol K$^2$) | $\beta$ (mJ/mol K$^4$) | $\delta$ (µJ/mol K$^6$) | $\theta_D$ (K) |
|---|---|---|---|---|
| LaAgAl$_3$ | 7 | 0.1029 | 1.1693 | 455 |
| CeAgAl$_3$ | 221 | 0.3590 | 0.0064 | 301 |
| PrAgAl$_3$ | 15 | 0.4855 | 0.0158 | 271 |

The estimated magnetic specific heat ($C_{4f}$) exhibits the broad maximum around 10 K for Ce and two maxima at 10 K and 30 K for Pr respectively, as shown in Fig.6. This could be associated with the Schottky contribution arising from CEF splitting of ground state multiplet. The solid lines shown in Fig.6 have been fitted to the magnetic heat capacity of CeAgAl$_3$ and PrAgAl$_3$ above ordering temperature based on the standard expression for Schottky heat capacity as given below,

$$C_{sch}(T) = \frac{R}{T^2}\left[\frac{\sum_i g_i e^{\frac{-E_i}{k_B T}} \sum_i g_i E_i^2 e^{\frac{-E_i}{k_B T}} - \left(\sum_i g_i E_i e^{\frac{-E_i}{k_B T}}\right)^2}{\left(\sum_i g_i e^{-E_i/k_B T}\right)^2}\right] \quad (3)$$

where $R$ is a gas constant, $E_i$ is the energy in units of temperature and $g_i$ is the degeneracy of the CEF split energy levels.

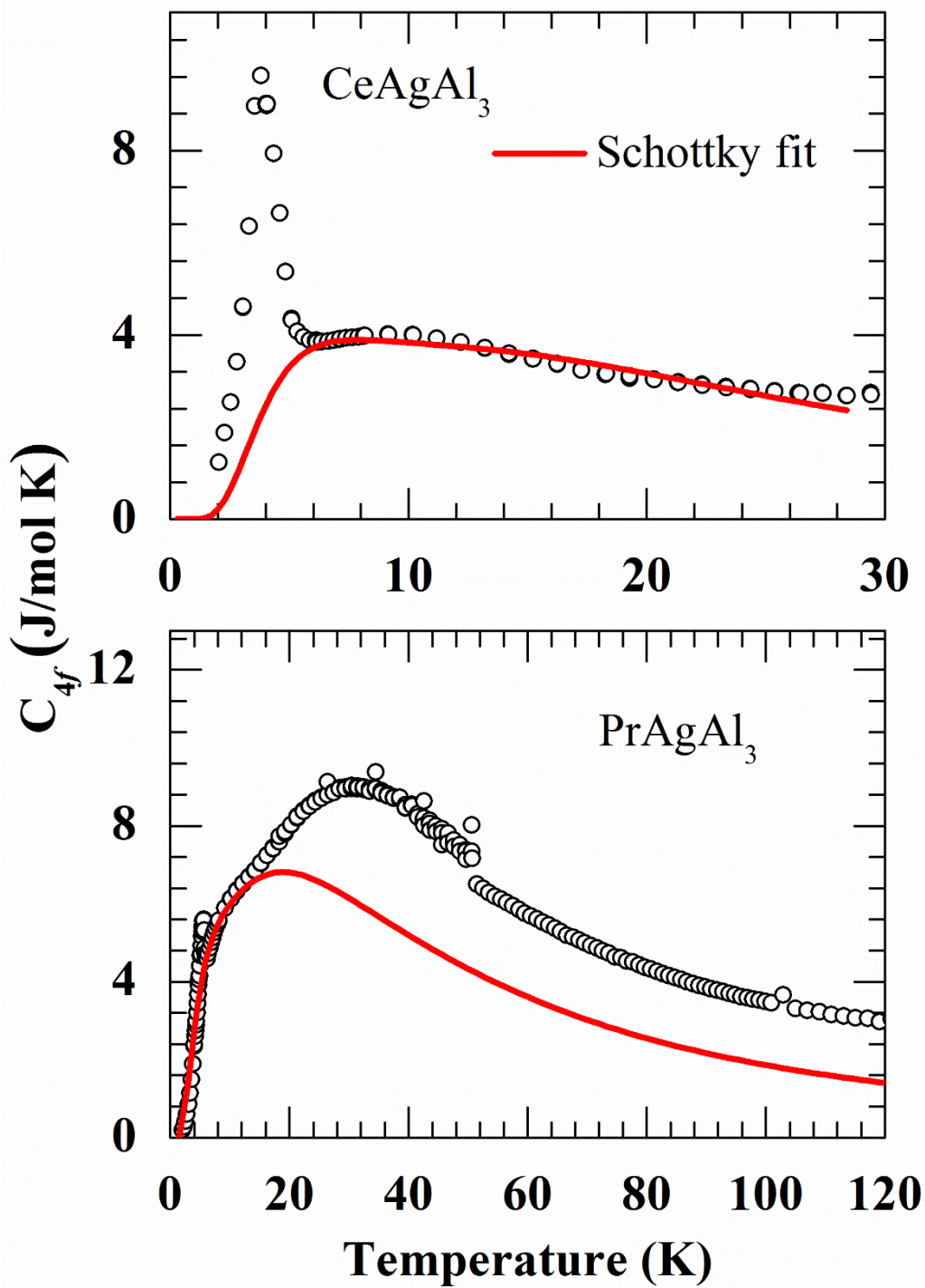

**Fig.6.** Temperature dependence of the 4$f$-derived specific heat ($C_{4f}$). In both figures, the solid lines represent the fits to the expression of Schottky heat capacity equation (3).

The results of the fits are shown as schematic energy level diagram in Fig. 7. It is well-known that, both Ce and Pr compounds are in the trivalent state, as confirmed from the magnetic measurements. In CeAgAl$_3$, the free ion Ce$^{3+}$ ($J$ = 5/2) have six-fold degenerate ground state multiplets, which split into three Kramer's doublets possessing the energies of first and second excited states as $\Gamma_1$ = 15 K and $\Gamma_2$ = 51 K, respectively. Pr$^{3+}$ is a non-Kramer's ion with $J$ = 4. In orthorhombic symmetry, under the CEF effect, 2$J$+1 degenerate ground state splits into nine singlet levels. The $C_{4f}(T)$ of Pr$^{3+}$ displays two broad maxima around 10 K and 30 K due to the short range correlations. The Schottky fit equation (3) to the maximum at (10 K) in $C_{4f}(T)$ gives the excitation energies as 15, 33, 53, 70, 90, 100, 160 and 180 K. Thus the ground and first excited levels are separated by an energy gap of 15 K, which could be taken as pseudo – doublet state that leads to magnetic ordering in PrAgAl$_3$ compound [15 -17].

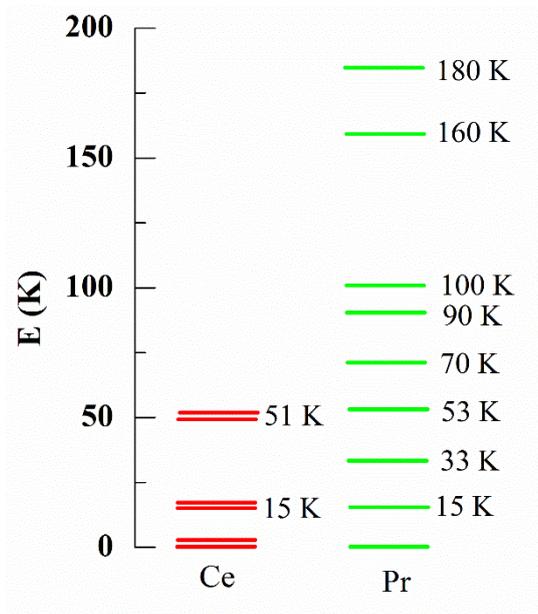

**Fig.7**. CEF energy levels from Schottky fit.

In CeAgAl$_3$, the application of magnetic field of 20 kOe shifts the magnetic ordering peak towards high temperatures with the reduction in peak height, as shown in Fig.8. PrAgAl$_3$ exhibits the similar behaviour of CeAgAl$_3$ (Fig.8). An external field of $H \geq 2.5$ kOe shifts the $T_C$ towards the higher temperature. For higher field of 90 kOe, the magnetic anomaly displays a clear broadening together with shifting to higher temperatures, which point to the ferromagnetic character of these compounds.

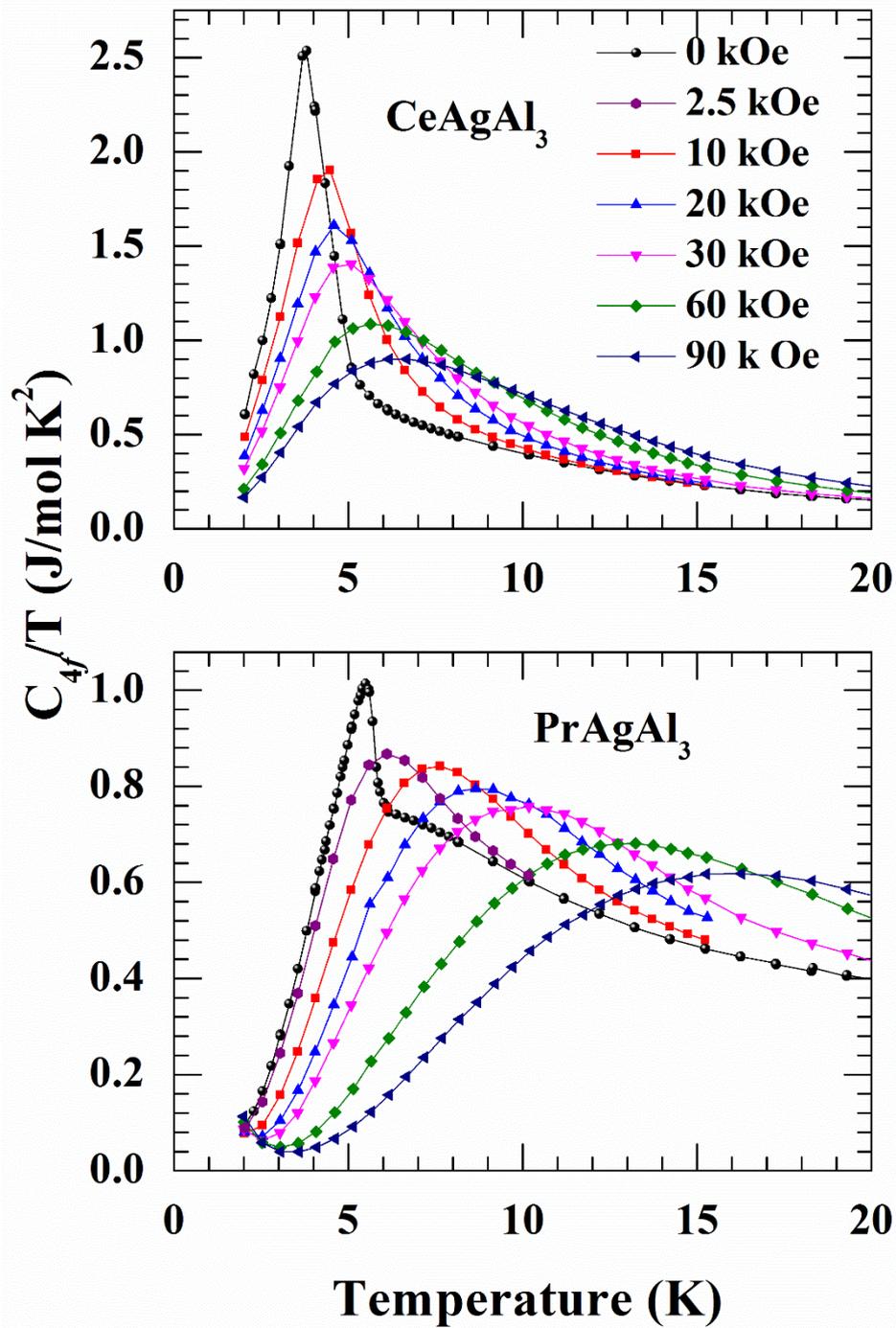

**Fig. 8.** The plots of $C_{4f}(T)/T$ versus T for RAgAl$_3$ (R = Ce and Pr) in various applied magnetic fields.

The $\gamma$ derived from fitting the equation (1) for CeAgAl$_3$ in the paramagnetic state ($T > T_C$) and magnetically ordered state ($T < T_C$) show a mass enhancement of $\gamma \approx 221$ and 352 mJ/mol K$^2$ respectively, as shown in Fig.5. These large $\gamma$ values in both regions infer that CeAgAl$_3$ is a heavy-Fermion compound. The application of magnetic field of H = 90 kOe suppresses the $\gamma$ to a moderate value of 139 mJ/mol K$^2$ at 2 K (see Fig.8). Hence, a moderate enhancement in the density of states is seen even after the application of high magnetic field 90 kOe. In the case of PrAgAl$_3$, the electronic coefficient $\gamma = 15$ mJ/mol K$^2$ is obtained in the paramagnetic state. C$_{4f}$(T) of PrAgAl$_3$, shows a slight upturn at 3K for an applied field of 90 kOe revealing the nuclear Schottky effect

## 5. Electrical resistivity and magnetoresistance (MR)

Fig.9. shows the temperature dependence of resistivity of RAgAl$_3$ (R = Ce and Pr), from 2 K to 300 K in zero field and in applied field of $H = 90$ kOe. The resistivity shows a metallic behaviour down to 1.8 K, which drops rapidly with a slope change around $T_C = 3.8$ K and 5.5 K for CeAgAl$_3$ and PrAgAl$_3$, respectively. This is due to the gradual freezing of spin disorder scattering. The observed small hump at 250 K in PrAgAl$_3$ may be due to the appearance of cracks in sample. At low temperature, the compounds do not follow the simple Fermi liquid (FL) behaviour due to the magnetic ordering. The residual resistivity ratios (RRR) for Ce and Pr have been determined from the relation $\rho(300\ K)/\rho_0(2\ K)$, where $\rho_0$ is residual resistivity at 2 K. We have obtained rather good residual resistivity ratios of 5.6 and 5.8 for Ce and Pr, respectively, which reflect the quality of the sample. The application of magnetic field of 90 kOe in CeAgAl$_3$ lowers the spin disorder scattering due to increase in the magnetic order, hence the slope tends to be smoothened around the ordering temperature. But in PrAgAl$_3$, Fermi liquid ($\rho \propto T^2$) behaviour is found in $\rho$ (T) for 90 kOe at low temperatures.

Fig.10. shows the *MR(H)* as a function of applied magnetic fields for two different temperatures $T = 2$ K and 6 K. The magnetoresistance (MR) can be defined as

$$MR(T,H) = \frac{\rho(T,\ H) - \rho(T,0)}{\rho(T,0)} 100\ \% \qquad (5)$$

Both the compounds CeAgAl$_3$ and PrAgAl$_3$ clearly display negative MR with the magnitudes of 19 % and 4 % respectively up to 90 kOe for 6 K ( near to the $T_C$), which is the usual case of ferromagnetic compound due to the suppression of spin-fluctuation at the domain walls. In the case of PrAgAl$_3$, there is an upturn in *MR(H)* beyond 20 kOe, which is not clear at the moment.

At low temperature (2 K) and high field, the magnitude of *MR (H)* is positive for both the compounds, which is associated with the Lorentz force effect as observed in metallic systems because the effect of the magnetic field is to increase the cyclotron frequency ($\omega_C$) of the electron which can be defined as $\omega_C \tau = \dfrac{H}{\rho_{H=0} n e}$, where τ is the collision time [18].

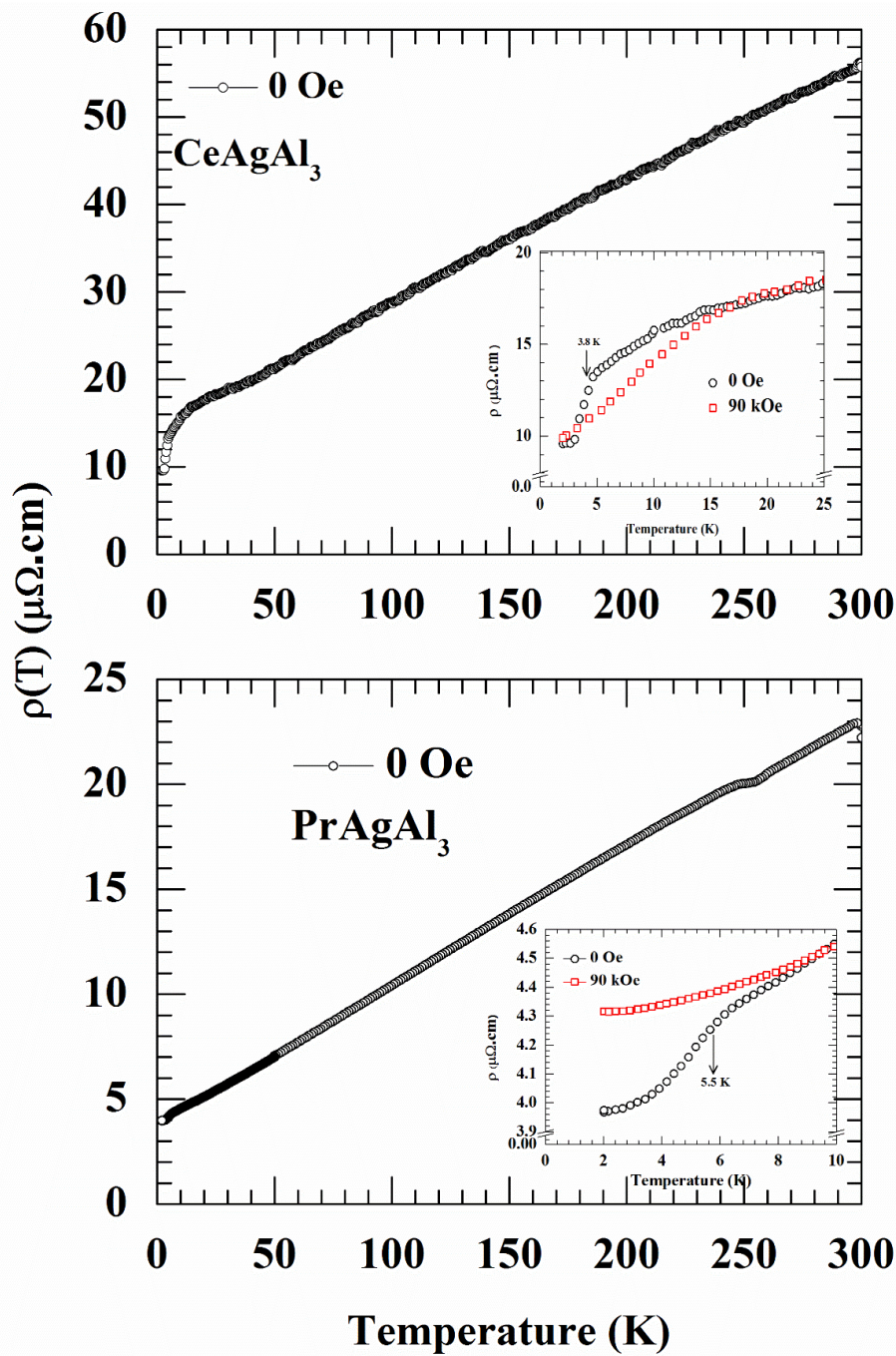

**Fig. 9**. Electrical resistivity ρ (*T*) of RAgAl$_3$ (R = Ce and Pr) measured in zero and 90 kOe field. Insets show the low temperature ρ (*T*) versus temperature.

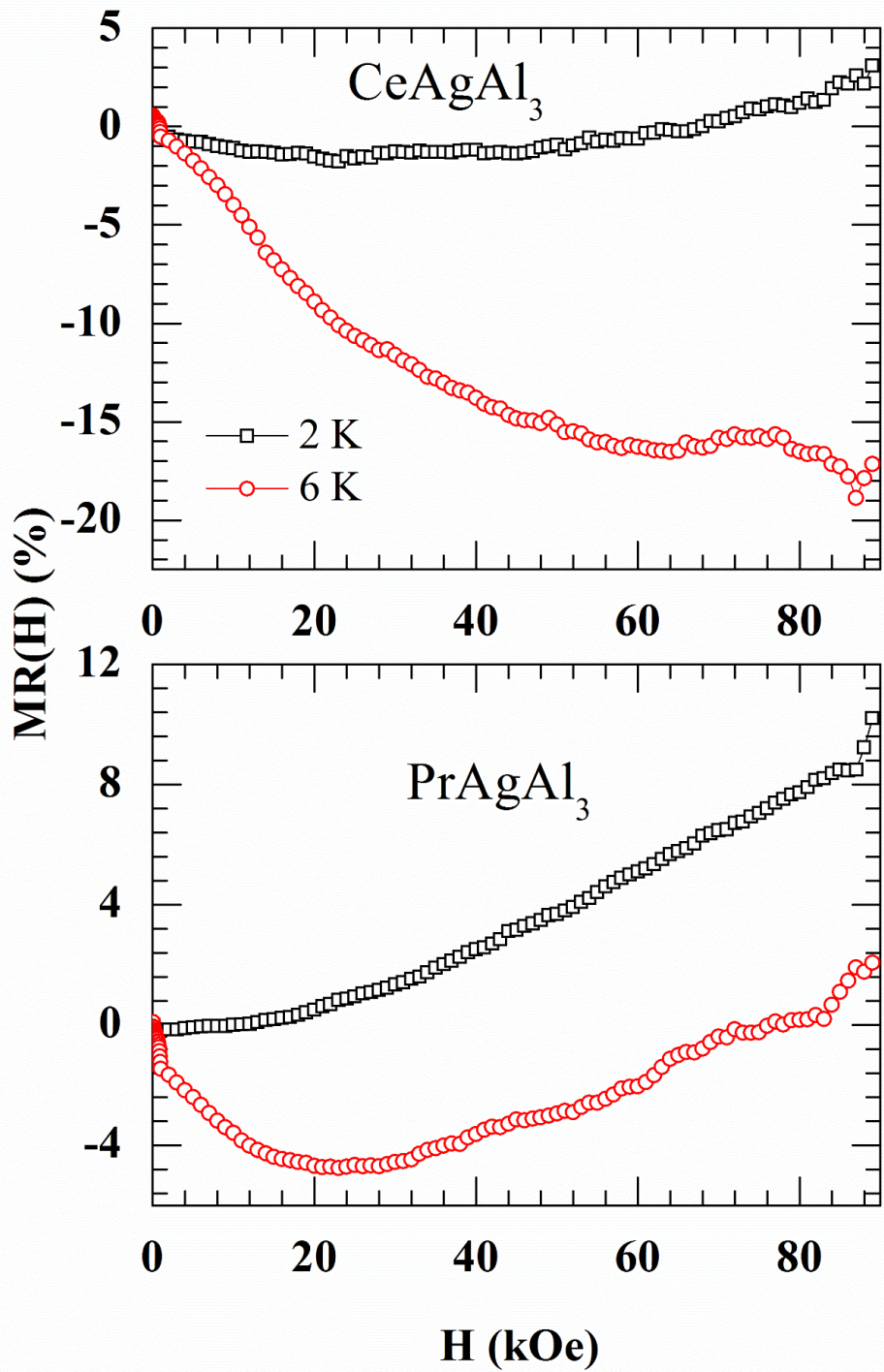

**Fig.10**. Magnetoresistance of RAgAl$_3$ (R = Ce and Pr) at 2 K and 6 K as function of magnetic fields up to 90 kOe

## 6. Conclusion

We have investigated the structure, magnetic, thermodynamic and transport properties of polycrystalline RAgAl$_3$ (R = La, Ce and Pr) compounds. The compounds of RAgAl$_3$ series (R = La, Ce and Pr) crystallize in the orthorhombic structure, which is distorted from tetragonal BaAl$_4$ structure. From different measurements, it has been concluded that Ce and Pr compounds undergo ferromagnetic ordering at low temperatures $T_C$ = 3.8 K and 5.5 K, respectively. The obtained effective magnetic moments were found to be close with trivalent Ce$^{3+}$ and Pr$^{3+}$ free-ion moments. ZFC and FC modes of dc magnetization along with the moderate frequency dependence in ac susceptibility of CeAgAl$_3$ indicates the domain wall pinning effect. CEF levels have been deduced from magnetic heat capacity using Schottky specific heat for both the compounds. From the fit it is understood that the ground and first excited states of PrAgAl$_3$ are separated by an energy difference of 15 K, which forms a pseudo-doublet ground state that leads to magnetic ordering in this compound. The large value of Sommerfeld coefficient before and after application of magnetic field suggest that CeAgAl$_3$ is a heavy Fermion compound. The electrical resistivity falls off rapidly below 3.8 K and 5.5 K for Ce and Pr, respectively around the ordering temperature, due to the loss in the spin-disorder scattering. MR study shows negative magnitudes at temperatures close to Tc and thereby displaying ferromagnetic ordering in both the compounds.

## Acknowledgements

The authors (R.N) thank Department of Atomic Energy (DAE), Board of Research in Nuclear Sciences (BRNS) Govt. of India for supporting this work under DAE Young Scientists Research Award (No: 2010/20/37P/BRNS/2513). The author (S.N) also thank BRNS for awarding JRF in the project.The paper is the result of the Project implementation: University Science Park TECHNICOM for Innovation Applications Supported by Knowledge Technology, ITMS: 26220220182, supported by the Research & Development Operational Programme funded by the ERDF.

**References**

[1] E. Bauer, G. Hilscher, H. Michor, Ch. Paul, E.W. Scheidt, A. Gribanov, Yu Seropegin, H. Noel, M. Sigrist, P. Rogl, Heavy Fermion Superconductivity and Magnetic Order in Noncentrosymmetric CePt$_3$Si, Phys. Rev. Lett. 92 (2004) 027003.

[2] N. Kimura, K. Ito, K. Saitoh, Y. Umeda, H. Aoki, and T. Terashima, Pressure-Induced Superconductivity in Noncentrosymmetric Heavy-Fermion CeRhSi$_3$, Phys. Rev. Lett. 95 (2005) 247004,

[3] N. Egetenmeyer, J. L. Gavilano, A. Maisuradze, S. Gerber, D. E. MacLaughlin, G. Seyfarth, D. Andreica, A. Desilets-Benoit, A. D. Bianchi, Ch. Baines, R. Khasanov, Z. Fisk, and M. Kenzelmann, Direct Observation of the Quantum Critical Point in Heavy Fermion CeRhSi$_3$, Phys. Rev. Lett. 108 (2012) 177204.

[4] Yu. N. Grin, K. Hiebl and P. Rogl, Magnetism and structural chemistry of ternary gallides RENi$_x$Ga$_{4-x}$(RE = La, Ce, Pr, Nd, Sm, Gd, Tb) and LaCo$_{0.5}$Ga$_{3.5}$, J. Less-Common Met, 162 (1990) 36l - 369.

[5] Yu. N. Grin, P. Rogl and K. Hiebl, Structural chemistry and magnetic behaviour of ternary gallides REPt$_x$Ga$_{4-x}$ (RE - La, Ce, Pr, Nd, Sm), J. Less-Common Met, 136 (1988) 329 - 333.

[6] E. V. Sampathkumaran and I. Das, Chemical pressure effects on the crystallographic and magnetic behavior of CeNiGa$_3$, Phys. Rev. B. 53 (1996) 8200.

[7] R. Nagalakshmi, R. Kulkarni, S. K. Dhar, A. Thamizhavel, V. Krishnakumar, M. Reiffers, I. Čurlík, H. Hagemann, D. Lovy and S. Nallamuthu, Magnetic properties of the tetragonal RCuGa$_3$ (R = Pr, Nd and Gd) single crystals, J. Magn. Magn. Mater. 386 (2015) 37–43.

[8] S. Nallamuthu, S. Selva Chandrasekaran, P. Murugan, M. Reiffers and R. Nagalakshmi, Magnetic, thermodynamic and transport properties of novel non-centrosymmetric RCoSi$_3$ (R = Pr, Nd and Sm) compounds, J. Magn. Magn. Mater, 416 (2016), 373–383.

[9] T. Muranaka and J. Akimitsu, Thermodynamic properties of ferromagnetic Ce-compound, CeAgAl$_3$, Physica C. 460–462 (2007) 688–690.

[10] C. Franz, A. Senyshyn, A. Regnat, C. Duvinage, R. Schönmann, A. Bauer, Y. Prots, L. Akselrud, V. Hlukhyy, V. Baran and C. Pfleiderer, Single crystal growth of CeTAl$_3$ (T = Cu, Ag, Au, Pd and Pt), J. Alloys Compd. 688, (2016) 978–986.

[11] A.C. Larson and R.B. Von Dreele, General Structure Analysis System (GSAS), Los Alamos National Laboratory Report LAUR, (2004) 86-748.

[12] S. Seidel, R. D. Hoffmann, R. Pottgen, SrPdGa$_3$ - An orthorhombic superstructure of the ThCr$_2$Si$_2$ type, Z. Krist. 229 (6) (2014) 421- 426.

[13] P. A. Joy, A. Kumar and S. K. Date, The relationship between field-cooled and zero-field-cooled susceptibilities of some ordered magnetic systems, J. Phys. Condens. Matter, 10 (1998) 11049–11054

[14] J. A. Mydosh, Spin Glass: An Experimental Introduction, Taylor & Francis, London, 1993.

[15] P. K. Das, A. Bhattacharyya, R. Kulkarni, S. K. Dhar, and A. Thamizhavel, Anisotropic magnetic properties and giant magnetocaloric effect of single-crystal PrSi, Phys. Rev. B. 89 (2014) 134418.

[16] L. D. Tung, D. M. Paul, M. R. Lees, P. Schobinger-Papamantellos, K.H.J. Buschow, Specific heat studies of PrCoAl$_4$ single crystal, J. Magn. Magn. Mater., 281 (2004) 378–381.

[17] M. Loewenhaupt, Crystal fields in low symmetry systems, Physica B. 163(1990) 479.

[18] R. Rawat, P. Kushwaha and I. Das, Magnetoresistance studies on RPd$_2$Si (R = Tb, Dy, Lu) compounds, J. Phys. Condens. Matter. 21 (2009) 306003.